\documentclass[10pt,twoside,twocolumn]{article}
\usepackage[super,sort&compress,comma]{natbib} 
\usepackage{times,mathptmx}
\usepackage{sectsty}
\usepackage{balance} 
\usepackage{bm,amsmath,amssymb}

\usepackage{graphicx}
\usepackage{lastpage}
\usepackage[format=plain,justification=raggedright,singlelinecheck=false,font=small,labelfont=bf,labelsep=space]{caption} 
\usepackage{fancyhdr}
\begin{document}

\thispagestyle{plain}
\fancypagestyle{plain}{
\renewcommand{\headrulewidth}{1pt}}
\renewcommand{\thefootnote}{\fnsymbol{footnote}}
\renewcommand\footnoterule{\vspace*{1pt}%
\hrule width 3.4in height 0.4pt \vspace*{5pt}} 
\setcounter{secnumdepth}{5}

\makeatletter 
\def\subsubsection{\@startsection{subsubsection}{3}{10pt}{-1.25ex plus -1ex minus -.1ex}{0ex plus 0ex}{\normalsize\bf}} 
\def\paragraph{\@startsection{paragraph}{4}{10pt}{-1.25ex plus -1ex minus -.1ex}{0ex plus 0ex}{\normalsize\textit}} 
\renewcommand\@biblabel[1]{#1}            
\renewcommand\@makefntext[1]%
{\noindent\makebox[0pt][r]{\@thefnmark\,}#1}
\makeatother 
\renewcommand{\figurename}{\small{Fig.}~}
\sectionfont{\large}
\subsectionfont{\normalsize} 

\fancyfoot{}
\fancyfoot[RO]{\footnotesize{\sffamily{1--\pageref{LastPage} ~\textbar  \hspace{2pt}\thepage}}}
\fancyfoot[LE]{\footnotesize{\sffamily{\thepage~\textbar\hspace{3.45cm} 1--\pageref{LastPage}}}}
\fancyhead{}
\renewcommand{\headrulewidth}{1pt} 
\renewcommand{\footrulewidth}{1pt}
\setlength{\arrayrulewidth}{1pt}
\setlength{\columnsep}{6.5mm}
\setlength\bibsep{1pt}

\twocolumn[
  \begin{@twocolumnfalse}
\noindent\LARGE{\textbf{Lipid membrane deformation in response to a local pH modification: theory and experiments}}
\vspace{0.6cm}

\noindent\large{\textbf{Anne-Florence Bitbol,\textit{$^{a}$} Nicolas Puff,\textit{$^{b a}$} Yuka Sakuma,\dag\textit{$^{c}$} Masayuki Imai,\dag\textit{$^{c}$} Jean-Baptiste Fournier,\textit{$^{a}$} and Miglena I. Angelova$^{\ast}$\textit{$^{b a}$}}}\vspace{0.5cm}
 
\vspace{0.6cm}

\noindent \normalsize{We study the deformation of a lipid membrane in response to a local pH modification. Experimentally, a basic solution is microinjected close to a giant unilamellar vesicle. A local deformation appears in the zone of the membrane that is closest to the micropipette, and relaxes when the injection is stopped. A theoretical description of this phenomenon is provided. It takes fully into account the spatiotemporal evolution of the concentration of hydroxide ions during and after the microinjection, as well as the linear dynamics of the membrane. This description applies to a local injection of any substance that reacts reversibly with the membrane lipids. We compare experimental data obtained in the domain of small deformations to the results of our linear description, and we obtain a good agreement between theory and experiments. In addition, we present direct experimental observations of the pH profile on the membrane during and after the microinjection, using pH-sensitive fluorescent lipids. }
\vspace{0.5cm}
 \end{@twocolumnfalse}
  ]

\footnotetext{\textit{$^{a}$~Universit\'e Paris Diderot, Paris 7, Sorbonne Paris Cit\'e, Laboratoire Mati\`ere et Syst\`emes Complexes (MSC), CNRS UMR 7057, B\^atiment Condorcet, Case courrier 7056, F-75205 Paris Cedex 13, France; E-mail: miglena.anguelova@upmc.fr}}
\footnotetext{\textit{$^{b}$~Universit\'e Pierre et Marie Curie (UPMC), Paris 6, Physics Department -- UFR 925, F-75005, Paris, France}}
\footnotetext{\textit{$^{c}$~Department of Physics, Ochanomizu University, Bunkyo, Tokyo 112-8610, Japan}}
\footnotetext{\dag~Present address: Department of Physics, Tohoku University, Aramaki, Aoba, Sendai 980-8578, Japan}

\section*{Introduction}
During cell life, membranes are subjected to an inhomogeneous and variable environment, which can be strongly coupled to biological processes. Local pH inhomogeneities at the cellular scale are ubiquitous and especially important. For instance, in various types of migrating cells, an actively generated intracellular pH gradient exists along the axis of movement, and it appears to be essential for cell migration~\cite{Martin11}. The effect of an extracellular pH gradient on cell migration and morphology has also been demonstrated in the case of human melanoma cells~\cite{Stuwe07}. Local pH is also of great importance in mitochondria. The $H^+$--ATP synthase enzymes that synthesize adenosine triphosphate (ATP), the cell's fuel, are powered by the local pH difference across the inner membrane of mitochondria. Both the proton pumps that maintain actively this pH difference and the enzymes that use it are located in dynamic membrane invaginations called cristae, where pH heterogeneities are thought to be especially important~\cite{Davies11}. 

In order to understand the fundamental phenomena at stake in the response of a biological membrane to a local pH modification, we work on giant unilamellar vesicles (GUVs). These biomimetic membranes are made of amphiphilic lipid molecules that self-assemble into closed bilayers in water. The local environment of a GUV can be modified in a controlled way by microinjecting a reagent close to it~\cite{Angelova99}. In particular, we showed previously that modifying locally the pH on a GUV through microinjection results into a local dynamic membrane deformation: the membrane shape is tightly coupled to local pH inhomogeneities~\cite{Khalifat08, Fournier09, Bitbol11_guv}. We also developed a theoretical description of the dynamics of a membrane subjected to a local modification affecting its physical properties, in the simple case of a constant modification of the membrane involving only one wavelength~\cite{Fournier09, Bitbol11_guv}. 

In this paper, we extend our theoretical description to account fully for the time-dependent profile of the fraction of chemically modified lipids in the membrane. This profile results from the diffusion of the basic solution in the water that surrounds the membrane. We solve analytically the corresponding diffusion problem using Green's functions, and we integrate numerically the linear equations describing the dynamics of the membrane for the resulting time-dependent profile of the fraction of chemically modified lipids in the membrane. We compare the results of this extended theoretical description to experimental measurements of the height of the membrane deformation during and after the local injection of a basic solution, in the regime of small deformations. In addition, we present a direct experimental visualization of the pH profile on the membrane during and after the microinjection, obtained using a pH-sensitive fluorescent membrane marker.

\section*{Materials and methods}
\subsection*{Membrane composition and vesicles preparation}

The following lipids, from Avanti Polar Lipids, were used without further purification: egg yolk L-$\alpha$-phosphatidylcholine (EYPC), brain L-$\alpha$-phosphatidylserine (PS), and the fluorescent lipid analog 1,2-dioleoyl-\textit{sn}-glycero-3-phosphoethanolamine-N-(carboxyfluorescein) ammonium salt (DOPE-CF). All other chemicals were of highest purity grade: calcein and NaOH, Sigma. 

Giant unilamellar vesicles were formed by the liposome electroformation method~\cite{Angelova86} in a thermostated chamber. Liposome preparations for phase contrast microscopy experiments were made with a unique lipid mixture of EYPC and PS with EYPC/PS 90:10 mol/mol. For fluorescence observations, 1\% (mol) of the fluorescent lipid analog DOPE-CF was added to this mixture. The particular electroformation protocol used in this work was the following: lipid mixture solutions were prepared in chloroform/diethyl ether/methanol (2:7:1) with a total lipid concentration of 1 mg/ml. A droplet of this lipid solution (1 $\mu$l) was deposited on each of the two parallel platinum wires constituting the electroformation electrodes, and dried under vacuum for 15 min. An AC electrical field, 10 Hz, 0.26 Vpp, was applied to the electrodes. Water (temperature $25^\circ\mathrm{C}$) was added to the working chamber (avoiding agitation). The voltage was gradually increased (for more than two hours) up to 1 Vpp and maintained during 15 more minutes, before switching the AC field off. The GUVs were then ready for further use. In each preparation at least 10 GUVs of diameter 50-80 $\mu$m were available.

Large unilamellar vesicles (LUVs) were prepared using the extrusion method~\cite{Extrusion91}, implemented as in Ref.~\cite{Khalifat11}. Samples were prepared by dissolving and mixing the above-mentioned lipids in chloroform/methanol (9.4:0.6 vol/vol) to obtain the desired composition (EYPC/PS 90:10 mol/mol, to which 1\% mol of DOPE-CF was then added). 

\subsection*{Microscopy imaging and micromanipulation}

We used a Zeiss Axiovert 200M microscope, equipped with a charged-coupled device camera (CoolSNAP HQ; Photometrics). The experiments were computer-controlled using the Metamorph software (Molecular Devices). The morphological transformations and the dynamics of the membrane were followed by phase contrast and fluorescence microscopy (Zeiss filter set 10, Ex/Em = 475/520 nm).

Tapered micropipettes for the local injection of NaOH were made from GDC-1 borosilicate capillaries (Narishige), pulled on a PC-10 pipette puller (Narishige). The inner diameter of the microcapillary used for performing the local injections onto a GUV was $0.3~\mu\mathrm{m}$. For these local injections, a microinjection system (Eppendorf femtojet) was used. The micropipettes were filled with basic solutions of NaOH with concentrations ranging from 5 to 100~mM. The injected volumes were on the order of picoliters and the injection pressure was 200 hPa. The positioning of the micropipettes was controlled by a high-graduation micromanipulator (MWO-202; Narishige). The injections were performed at different distances from the GUV surface, taking care to avoid any contact with the lipid membrane. 

\subsection*{Steady-state fluorescence measurements}

Steady-state fluorescence measurements of LUV samples were carried out with a Cary Eclipse spectrofluorimeter (Varian Instruments) equipped with a thermostated cuvette holder ($\pm 0.1^\circ\mathrm{C}$). Excitation and emission slits were adjusted to 5 nm. Fluorescence emission spectra were all recorded at $25^\circ\mathrm{C}$. All fluorescence measurements were carried out at a total lipid concentration of 0.2 mM. In the experiments, the pH of the LUV samples was gradually modified by adding aliquots of acid or basic solutions. The measurements were carried out after a few minutes of equilibration under agitation.

\section*{Experiments}
\subsection*{Observation of the membrane deformation} 

The chemical modification of the membrane was achieved by locally delivering a basic solution of NaOH close to the vesicle. This local increase of the pH should affect the head groups of the phospholipids PS and EYPC forming the external monolayer of the membrane~\cite{Bitbol11_guv}, as well as the fluorescent marker (when present). 

\begin{figure*}
  \centering
  \includegraphics[width=0.65\textwidth]{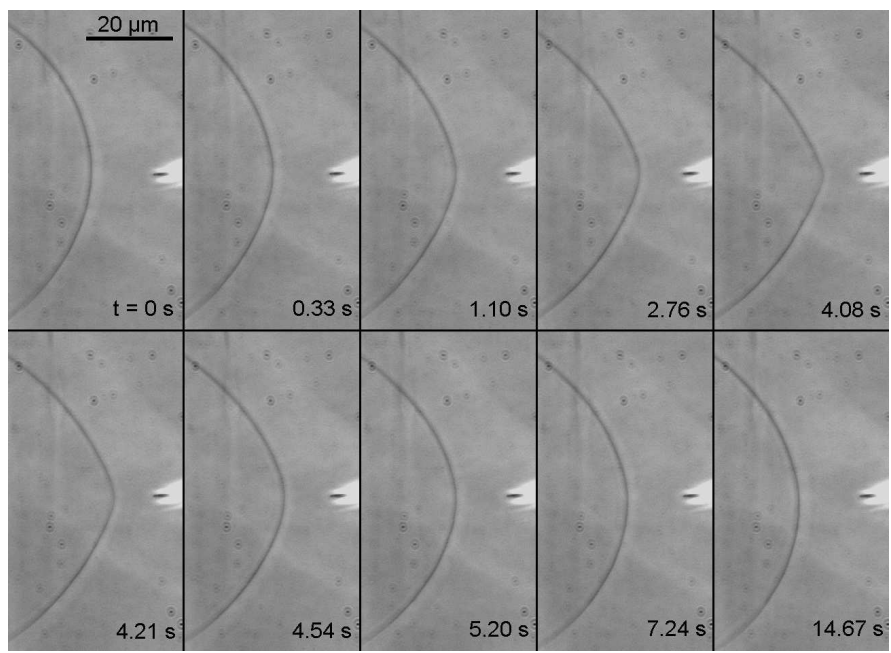}
  \caption{Typical microinjection experiment, lasting $T=4~\mathrm{s}$, observed using phase contrast microscopy. A local modulation of the pH on the vesicle membrane induces a smooth deformation of the vesicle (frames 0.33 to 4.08 s). The deformation is completely reversible when the NaOH delivery is stopped (frames 4.21 s to the end).
\label{Planche_Phase_Contrast}}
\end{figure*}

Figure~\ref{Planche_Phase_Contrast} shows a typical microinjection experiment. We inject the basic solution during a time $T=4~\mathrm{s}$. 
One can see in Fig.~\ref{Planche_Phase_Contrast} the vesicle before any microinjection (frame 0 s). A smooth local deformation of the vesicle develops toward the pipette during the microinjection (first line of images). Once the injection is stopped, the membrane deformation relaxes (second line of images). This deformation is fully reversible. For the sake of clarity, the deformation presented in Fig.~\ref{Planche_Phase_Contrast} is actually the largest this paper deals with. Indeed, we focus on the regime of small deformations in order to remain in the framework of our linear theory. In particular, it is necessary for our theory to be valid that the deformation height be much smaller than the distance between the membrane and the micropipette. 

\subsection*{Observation of the pH profile on the membrane}

We present here a direct experimental visualization of the pH profile on the membrane during and after the local microinjection of NaOH. For this, we use a pH-sensitive fluorescent membrane marker (DOPE-CF). The fluorescein group is a weak acid whose conjugate base has a strong fluorescence, and which is attached to phosphatidylethanolamine lipids.

\begin{figure}[h t b]
\centering
  \includegraphics[width=0.65\columnwidth]{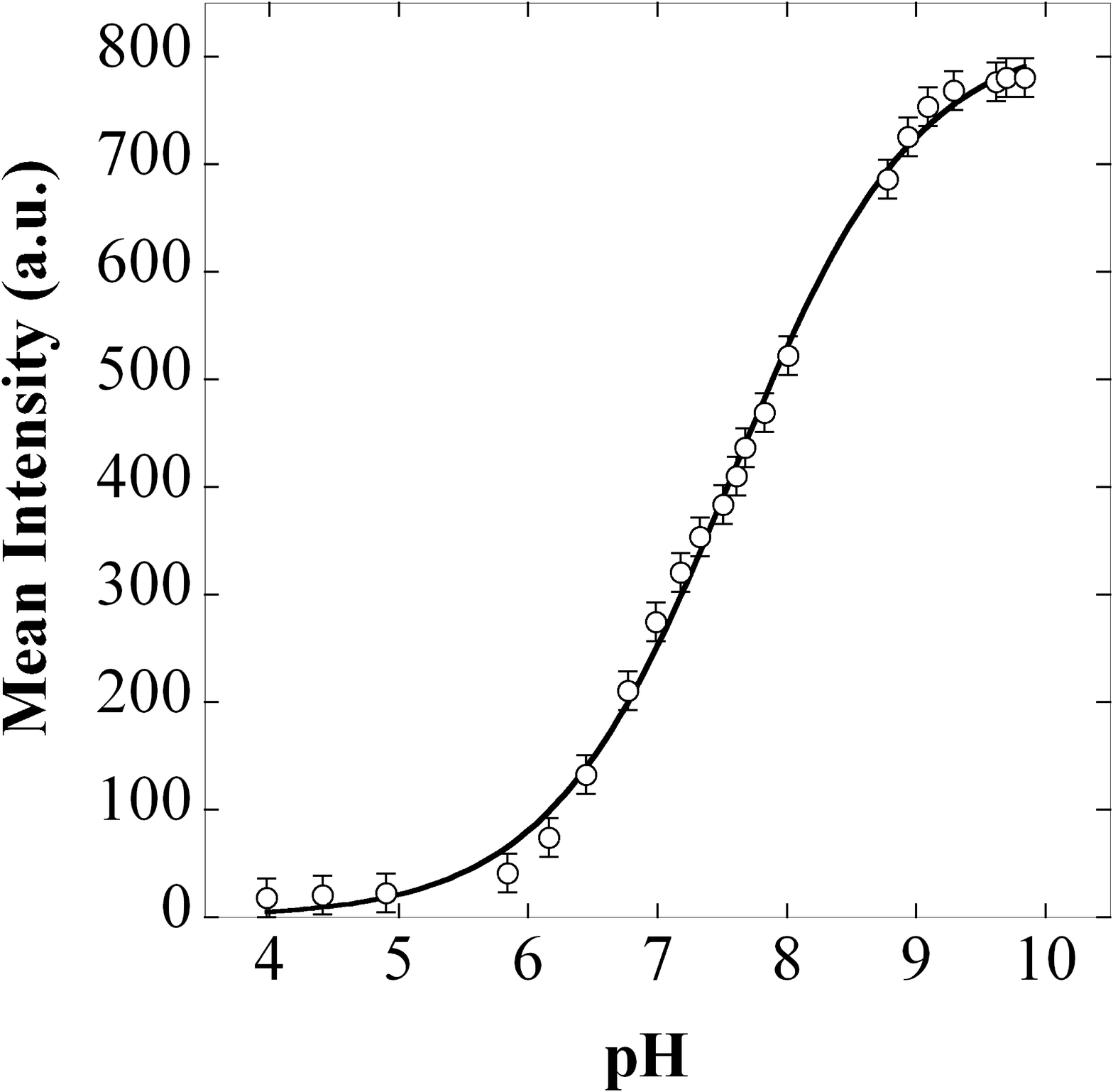}
  \caption{Mean (fluorescence) intensity as a function of the pH in a solution of LUVs whose membranes contain a pH-sensitive fluorescent marker (DOPE-CF). Dots: experimental data. Plain line: sigmoidal fit.}
  \label{I_pH}
\end{figure}

First of all, the dependence of the fluorescence on the pH was verified on a solution of LUVs (see Fig.~\ref{I_pH}). The dependence of the mean intensity of the LUV solution on the pH is well-described by the sigmoidal shape characteristic of an acid-base titration (see Fig.~\ref{I_pH}). The steep intensity rise is observed around pH 7.5, which makes this marker adequate to investigate pH increases starting from a neutral pH.

\begin{figure*}
  \centering
  \includegraphics[width=0.9\textwidth]{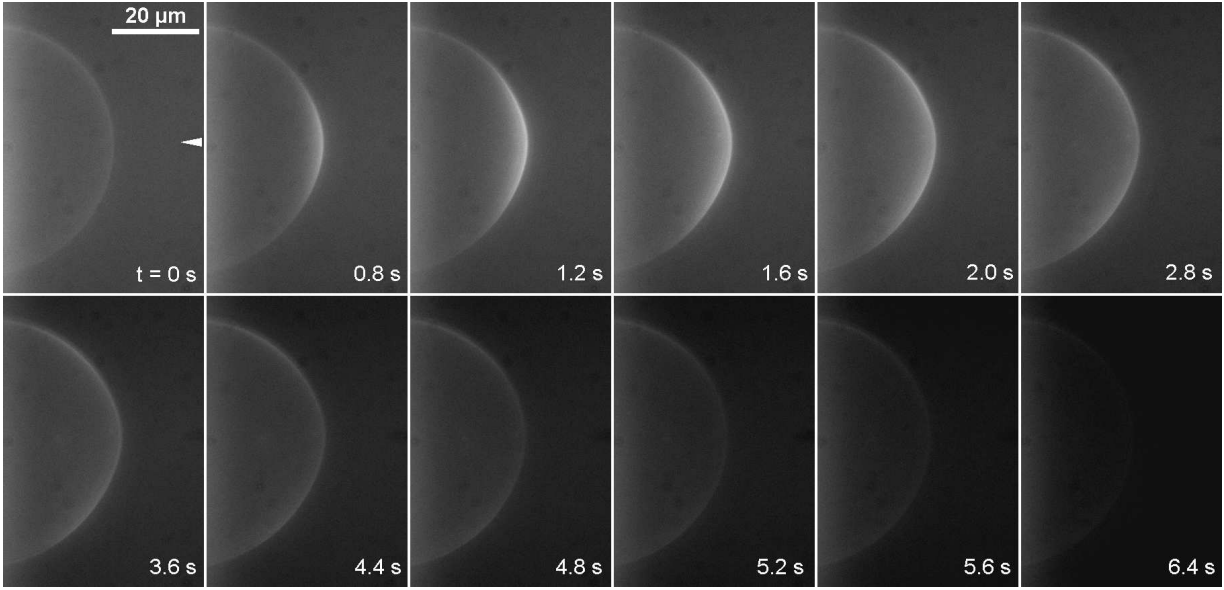}
  \caption{Typical microinjection experiment, lasting $T=4~\mathrm{s}$, observed using fluorescence microscopy. The white arrow represents the micropipette tip. The pH profile on the membrane is visualized directly together with the deformation of the membrane. The vesicle deforms progressively in response to the local pH increase (frames from 0 to 3.6 s). The increase of the intensity illustrates the pH increase. The deformation relaxes fully when the NaOH injection is stopped (frames from 4.4 s to the end). The initial fluorescence of the membrane at $t=0$, i.e., at pH 7, is weak but visible (see Fig.~\ref{I_pH}). The global darkening of the images with time is due to photobleaching.}
  \label{Planche_Fluo}
\end{figure*}

Using this fluorescent marker within a preparation of GUVs, it is possible to observe the pH profile on the membrane during and after the local microinjection of NaOH. Typical results are presented in Fig.~\ref{Planche_Fluo}. The pH profile on the membrane is visualized directly together with the deformation of the membrane. The vesicle deforms progressively in response to the local pH increase (frames from 0 to 3.6 s). The increase of the intensity in front of the micropipette and the lateral spreading of the bright zone during the injection are visible in the three first images. This illustrates the local pH increase on the membrane in front of the micropipette, and then on the sides, as the HO$^-$ ions diffuse from the micropipette tip towards the membrane. Later on, photobleaching occurs, and its effect is visible faster in the zone where the onset of fluorescence occurred sooner, i.e., just in front of the micropipette. The deformation relaxes fully when the NaOH injection is stopped (frames from 4.4 s to the end), and the fluorescence decreases at the same time, both due to the diffusion of the injected basic solution in the water surrounding the membrane after the end of the injection, and to photobleaching.

\begin{figure}[h t b]
\centering
  \includegraphics[width=0.9\columnwidth]{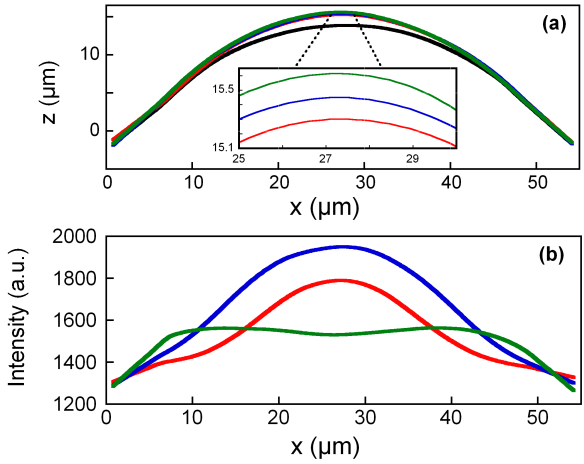}
  \caption{(a) Membrane shape profiles during the experiment presented in Fig.~\ref{Planche_Fluo} (black: t=0 s, red: t=0.8 s, blue: t=1.2 s, green: t=2.8 s). Insert: zoom of the central zone. (b) Intensity profiles on the membrane at the times 0.8, 1.2 and 2.8 s (same colors as in (a)).}
  \label{x_I}
\end{figure}

The membrane shape profiles and the membrane intensity profiles have been extracted from the images in Fig.~\ref{Planche_Fluo}, using the software Image J supplemented by our own plugins. The corresponding results are shown in Fig.~\ref{x_I} for the frames corresponding to times 0, 0.8, 1.2 and 2.8 s of Fig.~\ref{Planche_Fluo}. One can see first the local increase of intensity, and thus of pH, in front of the micropipette, as the membrane deforms (red curves), and the subsequent spreading of the high-pH zone while the pH increase continues (blue curves). In the last (green) curves, the effect of photobleaching is visible, as the central intensity decreases while the injection still continues and the membrane keeps deforming.

\section*{Theoretical description}

\subsection*{Membrane free energy and force density}
\begin{figure}[h t b]
\centering
  \includegraphics[width=0.9\columnwidth]{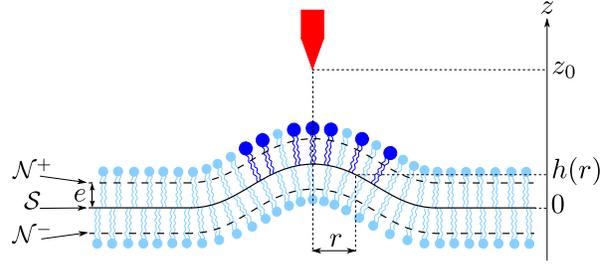}
  \caption{Sketch of the situation described (not to scale). Due to the injection of a basic solution from the micropipette (red) standing at $z_0$ above the membrane, some lipids are chemically modified (dark blue) in the external monolayer. The mass fraction of these modified lipids is denoted by $\phi$. The membrane deforms because of this local chemical modification: the shape of its midlayer $\mathcal{S}$ is described by $h(r)$. The variables $c=\nabla^2h$ and $r^\pm$ used in our theory are defined on $\mathcal{S}$, which is at a distance $e$ from the neutral surface $\mathcal{N}^\pm$ of monolayer $\pm$. }
  \label{Situation}
\end{figure}

In our experiments, only the external monolayer is directly affected by the chemical modification. Indeed, the permeation coefficient of hydroxide ions through a lipid membrane is $P_{\mathrm{OH}^-}\approx10^{-5}\,\mathrm{cm/s}$, which yields a negligible pH increase inside the vesicle on the timescale of our experiments~\cite{Elamrani83, Seigneuret86}. Thus, to study the membrane deformation, it is necessary to take fully into account the bilayer structure of the membrane. The free energy per unit area of a bilayer can be written as the sum of the free-energy densities of the two monolayers, which will be noted $f^+$ and $f^-$. We shall use a description of the membrane based on a local version of the area-difference elasticity model~\cite{Seifert93,Bitbol11_stress}. The local state of each monolayer is described by two variables: the total curvature $c$ defined on the membrane midlayer ($c$ is thus common to both monolayers), and the scaled density $r^\pm=(\rho^\pm-\rho_0)/\rho_0$ defined on the midlayer, where $\rho_0$ is a reference density. The free energy per unit area of a monolayer reads
\begin{equation}
f^\pm=\frac{\sigma_0}{2}+\frac{\kappa}{4}c^2\pm\frac{\kappa c_0}{2}c+\frac{k}{2}\left(r^\pm \pm ec\right)^2\,,
\label{fpm}
\end{equation}
where $\sigma_0$ is the membrane tension and $\kappa$ is its bending rigidity, while $k$ denotes the stretching modulus of a monolayer, and $e$ stands for the distance between the neutral surface~\cite{Safran} of each monolayer and the midsurface of the bilayer (see Fig.~\ref{Situation}). All these constants are identical for the two monolayers because they are assumed to be identical before the injection. The two monolayers also have opposite spontaneous curvatures, noted $\mp c_0$, since their lipids are oriented in opposite directions. We choose the sign convention for the curvature in such a way that a spherical vesicle has $c<0$. With this convention, the monolayer denoted ``$+$" in Eq.~(\ref{fpm}) is the external monolayer.

The expression (\ref{fpm}) of the free-energy density of a monolayer is a general second-order expansion around a reference state characterized by a flat shape ($c=0$) and a uniform density $\rho^\pm=\rho_0$ \cite{Bitbol11_stress}. It is valid for small deformations around this reference state: $r^\pm=\mathcal{O}(\epsilon)$ and $ec=\mathcal{O}(\epsilon)$, where $\epsilon$ is a small nondimensional parameter characterizing the deformation.

Let us call $\phi$ the mass fraction of the lipids of the external monolayer that are chemically modified by the local pH increase in the experiment (see Fig.~\ref{Situation}). Given the small concentrations used and the micropipette-membrane distance, the pH on the membrane remains well below the effective $\mathrm{pK_a}$ of the lipid head groups (which are of order 10 or higher~\cite{Tsui86, Lee99}). Hence, we assume that $\phi=\mathcal{O}(\epsilon)$. In addition, as justified above, we consider that the inner monolayer is not affected by the pH increase outside the vesicle. Including the small variable $\phi$ in addition to $r^+$ and $ec$, we can write to second order in $\epsilon$: ~\cite{Bitbol11_stress, Bitbol11_guv}
\begin{eqnarray}
f^+&=&\frac{\sigma_0}{2}+\sigma_1\phi+\frac{\sigma_2}{2}\phi^2+\tilde\sigma\left(1+r^+\right)\phi\ln\phi
+\frac{\kappa}{4}c^2\nonumber\\&&+\frac{\kappa}{2}\left(c_0+\tilde c_0\phi\right)c+\frac{k}{2}\left(r^++ ec\right)^2\,.
\label{fmod}
\end{eqnarray}
Note that the term in $\phi\ln\phi$ in Eq.~(\ref{fmod}) corresponds to mixing entropy~\cite{Bitbol11_stress}. The constants $\sigma_1$, $\sigma_2$, $\tilde\sigma$ and $\tilde c_0$ describe the response of the monolayer to $\phi$.

To first order, the chemical modification of the external monolayer (i.e., the presence of a nonzero $\phi$) causes both a change of the equilibrium density and a change of the spontaneous curvature of this monolayer~\cite{Bitbol11_guv}. On the neutral surface, these two effects are decoupled~\cite{Safran}: let us thus discuss them on this surface. For a homogeneous monolayer with constant mass, minimizing the free energy per unit mass $f^+/\rho^+$ with respect to $r^+$ and $c$ yields a spontaneous curvature changed by the amount $\delta c_0=\bar c_0\phi$ to first order, with $\bar c_0=\tilde c_0+2\sigma_1 e/\kappa$. This minimization also gives a scaled equilibrium density on the neutral surface changed by the amount $\delta r^+_\mathrm{eq}=\sigma_1\phi/k$ to first order.

The force density in each monolayer of a bilayer with lipid density and composition inhomogeneities described by the free-energy densities~(\ref{fpm})--(\ref{fmod}) has been derived in Ref.~\cite{Bitbol11_stress}. As we shall focus on small deformations of the membrane with respect to the plane shape, it is convenient to describe the membrane in the Monge gauge, i.e., by its height $z=h(x,y)$ with respect to a reference plane, $x$ and $y$ being Cartesian coordinates in the reference plane. Then, $c=\nabla^2 h+\mathcal{O}(\epsilon^3)$ for small deformations such that $\partial_i h=\mathcal{O}(\epsilon)$ and $\partial_i\partial_j h=\mathcal{O}(\epsilon)$ where $i,j\in\{x,y\}$. The force density in the membrane then reads to first order in $\epsilon$
\begin{align}
p_i^+(x,y)&=-k\,\partial_i\left(r^++e\nabla^2 h-\frac{\sigma_1}{k}\phi\right)\,,\label{pip_b}\\
p_i^-(x,y)&=-k\,\partial_i\left(r^--e\nabla^2 h\right)\,,\label{pim_b}\\
p_z(x,y)&=\sigma_0 \nabla^2 h-\tilde{\kappa}\nabla^4 h-k e\,\nabla^2 r_a-\left(\frac{\kappa \bar{c}_0}{2}-\sigma_1 e\right)\nabla^2\phi\,,
\label{pn_b}
\end{align}
where $p_i^\pm$ is the force density in monolayer ``$\pm$" acting in a direction $i$ tangential to the membrane, while $p_z=p_z^++p_z^-$ is the total normal force density in the membrane. In these formulas, we have introduced the antisymmetric scaled density $r_a=r^+-r^-$, and the constant $\tilde\kappa=\kappa+2ke^2$. 

Eq.~(\ref{pip_b}) shows that the equilibrium density change $\delta r^+_\mathrm{eq}=\sigma_1\phi/k$ yields a tangential force density proportional to $\bm{\nabla}\phi$. Besides, Eq.~(\ref{pn_b}) shows that both the spontaneous curvature change $\delta c_0=\bar c_0\phi$ and the equilibrium density change $\delta r^+_\mathrm{eq}$ yield a normal force density proportional to $\nabla^2\phi$ (see also Refs.~\cite{Bitbol11_stress,Bitbol11_guv}). Hence, both effects can drive a deformation of the membrane, but only the equilibrium density change can induce a significant tangential lipid flow.

\subsection*{Membrane dynamics}

The linear dynamical equations describing the joint evolution of the membrane shape $h$ and of the lipid densities $r^\pm$ were first developed in Refs.~\cite{Seifert93,Evans94}. We have adapted them to our study of the membrane deformation caused by a local chemical modification, first in the case where $\bar c_0=0$~\cite{Fournier09}, and then in the general case where both the equilibrium density and the spontaneous curvature of the external monolayer are changed due to the chemical modification~\cite{Bitbol11_guv}. In the present work, we use the same dynamical equations as in Ref.~\cite{Bitbol11_guv}, but we now take fully into account the spatiotemporal profile of $\phi$ during and after the microinjection, while Refs.~\cite{Fournier09, Bitbol11_guv} focused on the simple case of a constant, one-wavelength, $\phi$. 

Let us recall briefly the derivation of the linear dynamical equations describing the membrane dynamics~\cite{Seifert93,Bitbol11_guv}. One uses a normal force balance for the bilayer, which involves the normal elastic force density $p_z$ in Eq.~(\ref{pn_b}) and the normal viscous stress exerted on the membrane by the water that surrounds it. Besides, as each monolayer is a two-dimensional fluid, one writes down generalized Stokes equations involving the monolayer tangential elastic force densities in Eqs.~(\ref{pip_b}) or~(\ref{pim_b}), the two-dimensional viscous stress associated with the lipid flow, the tangential viscous stress exerted by the surrounding water, and the force density associated with the intermonolayer friction. The latter reads $\mp b\left(\bm{v}^+-\bm{v}^-\right)$, where $b$ is the intermonolayer friction coefficient~\cite{Evans94}, while $\bm{v}^\pm(x,y,t)$ is the two-dimensional velocity of the lipids within monolayer $\pm$. Since $\phi=\mathcal{O}(\epsilon)$, and since the lipid flow is induced by the chemical modification, one has $|\bm{v}^\pm|=\mathcal{O}(\epsilon)$. Finally, one uses mass conservation in each monolayer. Considering that each monolayer has a fixed total mass is justified here since the timescales of our experiments (about 10 seconds) are much shorter than the flip-flop characteristic time, which is assumed not to be significantly modified by the local chemical modification we study. 

The resulting equations can be expressed conveniently in terms of the in-plane Fourier transforms of the fields $h$, $r^\pm$ and $\phi$, denoted with a hat, e.g.,
\begin{equation}
\hat h(\bm{q},t)=\int_{\mathbb{R}^{2}}d\bm{r}\,h(\bm{r},t) e^{-i \bm{q}\cdot\bm{r}}\,,
\end{equation}
where $\bm{r}=(x,y)$ and $\bm{q}=(q_x,q_y)$ are two-dimensional vectors. The geometry of the problem we are studying features a cylindrical symmetry, the axis of revolution being that of the micropipette. Let us take the origin of the in-plane coordinates (i.e., $x$ and $y$) on the axis of the micropipette. Then, the fields $h$, $r^\pm$ and $\phi$ only depend on $r=\sqrt{x^2+y^2}$, and their Fourier transforms only depend on $q=\sqrt{q_x^2+q_y^2}$.

Combining the above-mentioned dynamical equations yields the following first-order linear differential equation on the two-dimensional variable $X(q,t)=(q\,\hat h(q,t),\,\hat r_a(q,t))$:~\cite{Bitbol11_guv} 
\begin{equation}
\frac{\partial X}{\partial t}(q,t)+M(q)\,X(q,t)=Y(q,t)\,,
\label{ED}
\end{equation}
where $t$ is time and $q$ is the norm of the two-dimensional wavevector. In Eq.~(\ref{ED}), we have introduced the matrix which describes the dynamical response of the membrane~\cite{Seifert93, Fournier09, Bitbol11_guv}:
\begin{equation}
M(q)=\left(\begin{array}{cc}
\displaystyle\frac{\sigma_0 q+\tilde\kappa q^3}{4\eta}
&-\displaystyle\frac{keq^2}{4\eta}\\\\
-\displaystyle\frac{keq^3}{b}
&\displaystyle\frac{kq^2}{2b}
\end{array}\right),\,\,
\label{ED_defs}
\end{equation}
where $\eta$ is the viscosity of water.
We have assumed that $\eta_2q^2\ll b$ and $\eta q\ll b$, where $\eta_2$ is the two-dimensional viscosity of the lipids, which is very well verified for all the wavevectors with significant weight in $\hat\phi$. Indeed, in the experimental situation we wish to describe, the modified lipid mass fraction $\phi$ has a smooth profile with a characteristic width larger than 1 $\mu$m, and typical values of the dynamical parameters, used throughout, are $\eta_2=10^{-9}\,\mathrm{J\,s/m^2}$, $b=10^9\,\mathrm{J\,s/m^4}$ (see Refs.~\cite{Pott02,Shkulipa06}), and $\eta=10^{-3}\,\mathrm{J\,s/m^3}$ for water. Besides, the forcing term in Eq.~(\ref{ED}) reads~\cite{Bitbol11_guv}:
\begin{equation}
Y(q,t)=\left(\begin{array}{c}
\displaystyle\frac{\kappa \tilde c_0 q^2}{8\eta}\hat\phi(q,t)\\\\
\displaystyle\frac{\sigma_1 q^2}{2 b}\hat\phi(q,t)
\end{array}\right).
\label{ED_defs_2}
\end{equation}

Eqs.~(\ref{ED}-\ref{ED_defs}) show that the antisymmetric density is coupled to the height of the membrane deformation. Physically, it is the symmetry breaking between the monolayers that causes the deformation of the membrane. In the present case, this symmetry breaking is caused by the chemical modification of certain membrane lipids in the external monolayer, i.e., to the presence of $\phi$. And indeed, Eq.~(\ref{ED_defs_2}) shows that the forcing term in Eq.~(\ref{ED}) is proportional to $\hat\phi (q,t)$.

The present theoretical description is general and applies to any local chemical modification of the membrane. Indeed, we have not made any specific assumption on $\phi$, apart that it is small. What changes with the nature of the reagent injected is the value of the constants $\sigma_1$ and $\bar c_0$, which describe the linear response of the membrane to the chemical modification.

\subsection*{Calculation of $\hat \phi(q,t)$}

The time evolution of the membrane deformation during and after the microinjection can be determined by solving Eq.~(\ref{ED}). To this end, we first need to determine $\hat \phi (q,t)$, which is involved in the forcing term $Y(q,t)$ (see Eqs.~(\ref{ED}--\ref{ED_defs})). The acid-base (and complexation) reactions that occur between the lipids and the injected hydroxide ions are reversible and diffusion-controlled~\cite{Eigen64}. Hence, the local mass fraction $\phi(r,t)$ of chemically modified lipids is determined by an instantaneous equilibrium with the hydroxide ions that are above the membrane. Let us denote by $c\left(r,z,t\right)$ the local concentration in hydroxide ions in the water that surrounds the vesicle, and let us assume that the unperturbed flat membrane is in the plane $z=0$, and that the micropipette stands at $z_0>0$ above it (see Fig.~\ref{Situation}). In the experimental conditions, the pH on the membrane remains well below the effective pKa of the lipids. Thus, the mass fraction of chemically modified lipids after the reaction is proportional to the concentration of HO$^-$ ions just above the membrane: $\phi(r,t)\propto c(r,z=0,t)$. In this linear regime, calculating $\phi(r,t)$ amounts to evaluating $c(r,z=0,t)$. 

\begin{figure*}
  \centering
  \includegraphics[width=0.9\textwidth]{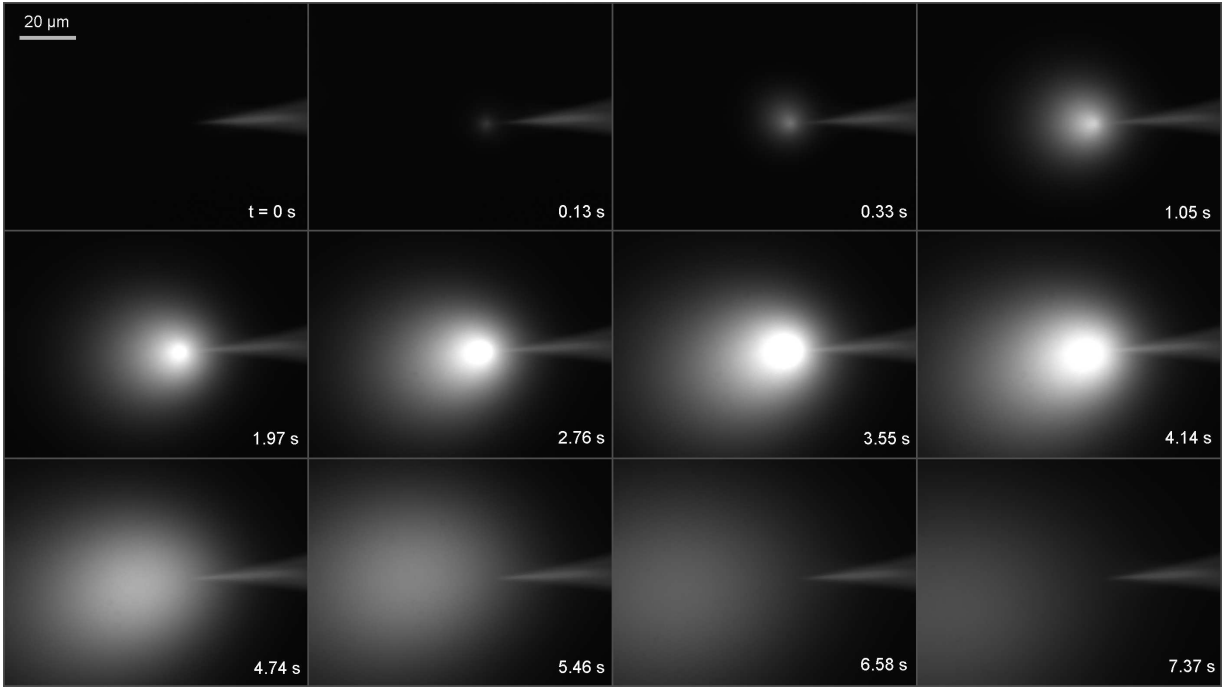}
  \caption{Injection of a water solution containing a fluorescent marker (calcein 96~mM) from a micropipette in the conditions under which the basic solution is injected in our experiments: injection pressure $\Delta P=200$ hPa, microcapillary inner diameter $0.3~\mu\mathrm{m}$, injection duration $T=4~\mathrm{s}$.}
  \label{Planche_Inj}
\end{figure*}

In our experiments, the NaOH solution is injected from a micropipette of inner diameter $d=0.3\,\mu\mathrm{m}$ and length $L\simeq 2$ cm, with an injection pressure $\Delta P=200$ hPa. Hence, we can estimate the average velocity $v_0$ of the NaOH solution when it just gets out the pipette, treating the flow in the micropipette as a Poiseuille flow: $v_0=\Delta P\,d^2/(32 \eta L)\simeq2\,\mu\mathrm{m.s}^{-1}$. Thus, given the small lengthscales and velocities involved in the microinjection, the P\'eclet number is very small: $Pe=v_0z_0/D\approx 10^{-2}$, where the order of magnitude of $z_0$ is $10\,\mu\mathrm{m}$, while $D$ is the diffusion coefficient of sodium hydroxide in water, which is given by $D=2/(1/D_{\mathrm{OH}^-}+1/D_{\mathrm{Na}^+})=2125\,\mu\mathrm{m^2/s}$ at infinite dilution and at $20 \,^{\circ}\mathrm{C}$~\cite{Cussler}. Hence, once the NaOH solution is out of the pipette, its dynamics is dominated by diffusion. This fact can also be verified experimentally by injecting a fluorescent substance in the conditions under which the basic solution is injected in our experiments. The observed quasi-spherical fluorescent ``cloud'', presented in Fig.~\ref{Planche_Inj}, illustrates well the dominant effect of diffusion.

Thus, we can consider that the NaOH solution simply diffuses from the micropipette tip $(r=0,z_0)$. Besides, since the membrane is a surface and $\phi\ll1$, it is possible to neglect the hydroxide ions that react with the membrane when calculating $c$. Given these simplifications, $c$ can be obtained by solving the (three-dimensional) diffusion equation 
\begin{equation}
\partial_t c-D\nabla^2 c=S\,,
\end{equation}
with the source term 
\begin{equation}
S(\bm{r},z,t)=S_0 \,\delta(\bm{r})\,\delta(z-z_0)\,\bm{1}_{[0,T]}(t)\,,
\end{equation}
where $\bm{1}_{[0,T]}$ is the indicator function of the interval $[0,T]$. This source term corresponds to a constant injection flow from the point $(r=0,z_0)$ for $0<t<T$, where we have chosen the injection start as our time origin, and we have called $T$ the time when the injection stops. In addition, the membrane imposes a Neumann boundary condition to $c$: $\partial_z c\left(r,z=h(r,t),t\right)=0$. Since we are working at first order in $\epsilon$, we simply use
\begin{equation}
\partial_z c\left(r,z=0,t\right)=0\,.
\label{nbc}
\end{equation}

The solution to this diffusion equation with the boundary condition Eq.~(\ref{nbc}) can be written as:
\begin{align}
c\left(r,z,t\right)&=\int_0^{t} \!\!dt' \int_{\mathbb{R}^2} \!\!\!\!d\bm{r'} \int_0^{+\infty}\!\!\!\!\!\!\!\!dz'\,S\left(\bm{r'},z',t'\right)\,G\left(|\bm{r}-\bm{r'}|,z,z',t-t'\right)\nonumber\\
&=S_0\int_0^{\mathrm{min}(t,\,T)} dt'\,G\left(r,z,z_0,t-t'\right)\,,
\label{cint}
\end{align}
where $G$ is the causal Green function of the diffusion equation which verifies the Neumann boundary condition Eq.~(\ref{nbc}). It can be expressed simply using the method of images~\cite{Alastuey}:
\begin{equation}
G\left(r,z,z',t\right)=G_\infty \left(r,z-z',t\right)+G_\infty \left(r,z+z',t\right)\,,
\label{G1}
\end{equation}
where we have introduced the infinite-volume causal Green function of the diffusion equation
\begin{equation}
G_\infty \left(r,z,t\right)=\frac{\theta(t)}{\left(4\,\pi\,D\,t\right)^{3/2}}\,\exp\left(-\frac{r^2+z^2}{4\,D\,t}\right)\,,
\label{G2}
\end{equation}
where $\theta$ denotes Heaviside's function. Using Eqs.~(\ref{cint}, \ref{G1}, \ref{G2}), it is straightforward to obtain an analytical expression for $c(r,z,t)$, and for its in-plane Fourier transform $\hat c (q,z,t)$. Since $\hat\phi(q,t)$ is proportional to $\hat c\left(q,z=0,t\right)$, we also obtain its analytical expression, which reads
\begin{equation}
\hat\phi\left(q,t\right)= \hat\phi_1\left(q,t\right)-\theta(t-T)\,\hat\phi_1\left(q,t-T\right)\,,
\label{phiq1}
\end{equation}
with
\begin{equation}
\hat\phi_1\left(q,t\right)\propto\mathrm{erf}\left(q\sqrt{Dt}-\frac{z_0}{2\sqrt{Dt}}\right) \frac{\cosh\left(qz_0\right)}{qz_0}- \frac{\sinh\left(qz_0\right)}{qz_0},
\label{phiq2}
\end{equation}
where erf denotes the error function. 
The value of the proportionality constant in this expression is not crucial for our study since all our dynamical equations are linear: it only affects the deformation and the antisymmetric scaled density by a multiplicative constant.

\subsection*{Resolution of the dynamical equations}

Now that we have determined $\hat \phi (q,t)$, which is involved in the forcing term $Y(q,t)$ of the dynamical equations (see Eqs.~(\ref{ED}--\ref{ED_defs})), the time evolution of the membrane deformation during and after the microinjection can be determined by solving the differential equation Eq.~(\ref{ED}). This can be done thanks to the method of variation of parameters. The square matrix $M$ defined in Eq.~(\ref{ED_defs}) has two real positive and distinct eigenvalues for all $q>0$.
Let us call these eigenvalues $\gamma_1$ and $\gamma_2$, and let us introduce the associated eigenvectors $V_1=(v_1,w_1)$ and $V_2=(v_2,w_2)$. For the initial condition $X(q,t=0)=(0,0)$, corresponding to a non-perturbed membrane (i.e., flat and with identical density in the two monolayers), we can write:
\begin{equation}
q\,\hat h(q,t)=\int_0^t ds\,\left[v_1\,e^{-\gamma_1\,t}A(s)+v_2\,e^{-\gamma_2\,t}\,B(s)\right],
\label{qh}
\end{equation}
where $A(t)$ and $B(t)$ are the solutions of the linear system
\begin{equation}
V_1\,A(t)\,e^{-\gamma_1\,t}+V_2\,B(t)\,e^{-\gamma_2\,t}=Y(q,t)\,.
\label{linsys}
\end{equation}

Then, in order to obtain the membrane deformation profile at time $t$, we perform an inverse Fourier transform: 
\begin{equation}
h(r,t)=\frac{1}{2\pi}\int_0^{+\infty} dq\,\left[q\,\hat h(q,t)\right] J_0(q\,r)\,,
\end{equation}
where we have used the cylindrical symmetry of the problem, introducing the Bessel function of the first kind and of zero order $J_0$. Thus, using Eq.~(\ref{qh}), we finally obtain
\begin{equation}
h(r,t)=\frac{1}{2\pi}\int_0^{+\infty}\!\!\!\!\!\! dq\,J_0(q\,r)\int_0^t ds\,\left[v_1\,e^{-\gamma_1\,t}A(s)+v_2\,e^{-\gamma_2\,t}\,B(s)\right].
\label{solution}
\end{equation}
Hence, we can obtain the spatiotemporal evolution of the membrane deformation during and after the microinjection by carrying out the integrals in Eq.~(\ref{solution}) numerically.

\section*{Comparison between theory and experiments}

In the previous section, we presented a theoretical description of the membrane deformation in response to the microinjection of a basic solution close to a membrane. In order to compare the predictions of this description to experimental results, we measured the height $H(t)=h(r=0,t)$ of the membrane deformation in front of the micropipette during the microinjection experiments described in the experimental section (see, e.g., Fig.~\ref{Planche_Phase_Contrast}). We carried out several microinjection experiments on different GUVs, all with an injection lasting $T=4$~s. For each GUV, several such experiments were conducted, with various distances $z_0$ between the micropipette and the membrane. 

Given that our theory is linear, we expect it to be valid in the regime where $\phi\ll 1$ and $H\ll z_0$. Besides, as the unperturbed membrane is considered flat in our theory, its domain of validity is restricted to $z_0\ll R$, where $R$ is the radius of the GUV. Hence, we strived to remain in these conditions, and we shall present here only the experimental results that match these conditions best. In practice, the radii of our largest GUVs were of order 60 to 80~$\mu$m. We thus focused on values of $z_0$ in the range 10 to 30~$\mu$m. Besides, we adjusted the HO$^-$ concentration for the various $z_0$ in order to obtain small but observable deformations, of order 1 to 5~$\mu$m. As long as we remain in the linear regime, the absolute value of the concentration of the injected solution only affects the deformation by a global proportionality constant. This was checked experimentally, albeit in a rough fashion given the uncertainty on the radii of the different micropipettes used. 

In order to compare the experimental data on $H(t)$ during and after the microinjection to the solutions of our theoretical equations, we normalize our experimental data on $H(t)$ by the value of $H(t=4\,\mathrm{s})$, corresponding to the end of the injection, for each experiment. This eliminates the effect of the unknown proportionality constant in our theoretical expression of $\hat \phi$ (see Eqs.~(\ref{phiq1}--\ref{phiq2})), as well as the experimental effect of the different concentrations and of the slightly different micropipette diameters. 

The membrane deformation $H(t)=h(r=0,t)$ predicted theoretically using Eq.~(\ref{solution}) together with Eq.~(\ref{linsys}) and Eqs.~(\ref{phiq1}--\ref{phiq2}) was calculated numerically. This was done in the case of four-second injections, with the values of $z_0$ corresponding to the different experiments, and taking typical values of the membrane constitutive constants: $\kappa=10^{-19}\,\mathrm{J}$, $k=0.1\,\mathrm{N/m}$, $b=10^9\,\mathrm{J.s.m}^{-4}$ and $e=1$~nm, and $\eta=10^{-3}\,\mathrm{J.s.m}^{-3}$ for the viscosity of water. 

In order to solve the dynamical equations, it was also necessary to assign a value to the parameter $\alpha=-ke\bar c_0/\sigma_1$, which quantifies the importance of the change of the spontaneous curvature relative to the change of the equilibrium density of the external monolayer as a result of the chemical modification. This parameter cannot be determined from an analysis of static and global modifications of the environment of the vesicle. Indeed, the area-difference elasticity model predicts that the equilibrium shape of a vesicle is determined by a combined quantity which involves both the equilibrium density and the spontaneous curvature~\cite{Lee99, Miao94}. A rough microscopic lipid model based on geometry yields $\alpha\approx 1$~\cite{Bitbol11_guv}: the effect of the change of the spontaneous curvature and that of the equilibrium density should have the same order of magnitude. In the absence of any experimental measurement of $\alpha$, we took $\alpha=1$ in our calculations. We also checked that the agreement between theory and experiment was not as good for $\alpha=0.1$ and $\alpha=10$ as it is for $\alpha=1$.

\begin{figure}[h t b]
\centering
  \includegraphics[width=\columnwidth]{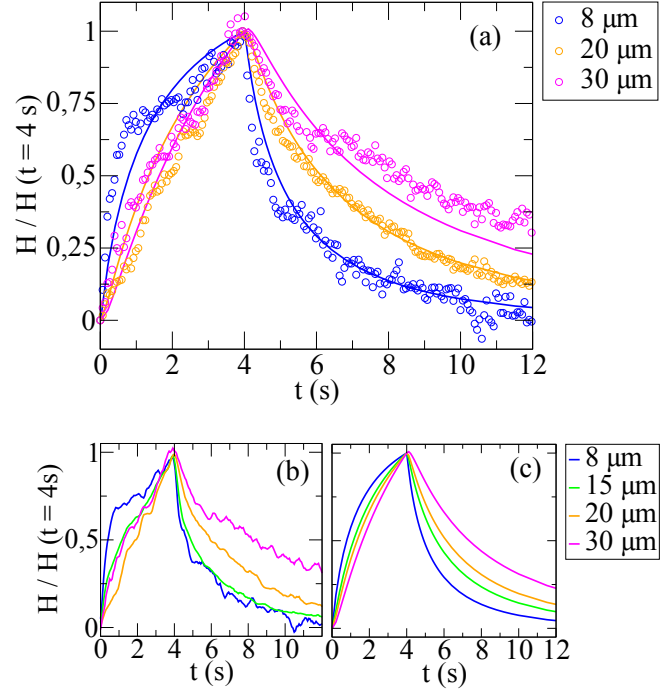}
  \caption{Normalized height $H(t)/H (t=4~\mathrm{s})$ of the membrane deformation, during and after microinjections lasting $T=4$~s. The microinjections were carried out on the same GUV at different distances $z_0$, corresponding to the different colors. (a) Comparison between experimental data and theoretical calculations for three experiments. Dots: raw experimental data (one data point was taken every 57~ms). Plain lines: numerical integration of our dynamical equations with $\sigma_0=1.5 \times 10^{-8}$~N/m. (b) Full set of experimental data; a moving average over 4 successive points was performed to reduce the noise. (c) Full set of numerical data for values of $z_0$ corresponding to those of the experiments.}
  \label{111102}
\end{figure}

\begin{figure}[h t b]
\centering
  \includegraphics[width=\columnwidth]{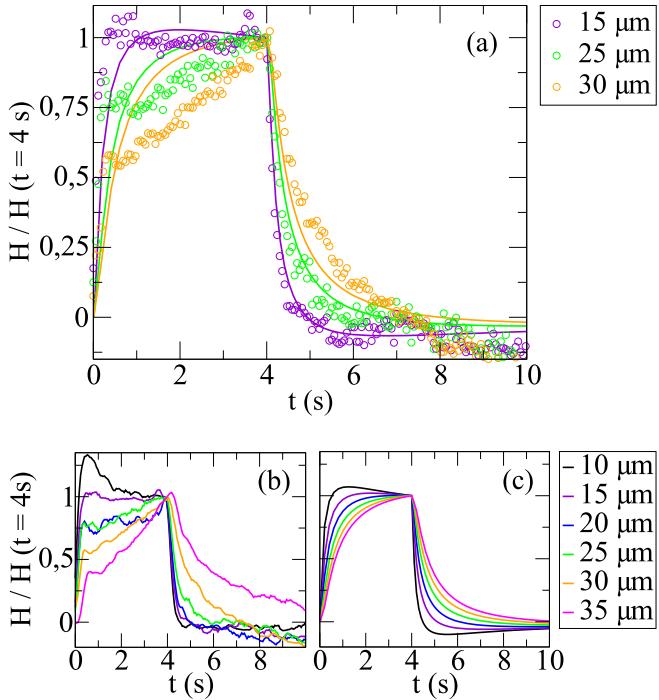}
  \caption{Similar data as on Fig.~\ref{111102}, but for another, less flaccid, vesicle. Here, the numerical integration of our dynamical equations was carried out for $\sigma_0=5 \times 10^{-7}$~N/m.}
  \label{111024}
\end{figure}

Figs.~\ref{111102} and \ref{111024} show the results obtained for several values of $z_0$, on two different vesicles. As these vesicles have the same lipid composition, their membranes share the same constitutive constants, as well as the same constants $\sigma_1$ and $\bar c_0$ that describe their response to $\phi$. On the contrary, the membrane tension $\sigma_0$ is highly variable among vesicles. We did not measure this tension during the experiments, but since the vesicles are flaccid, $\sigma_0$ should be in the range $10^{-8}-10^{-6}$~N/m. It was visible that the vesicle corresponding to Fig.~\ref{111102} was more flaccid than that corresponding to Fig.~\ref{111024}. The parameter $\sigma_0$ was adjusted in these two cases, with all the other parameters kept constant at the above-mentioned values. More precisely, we integrated numerically our dynamical equations assuming various values of the tension $\sigma_0$. For each of the two vesicles, we calculated the total chi-square between these numerical results and the three experimental data sets that respected best the hypotheses of our theory, i.e., small deformation and stable injection pressure (see Figs.~\ref{111102}(a) and \ref{111024}(a)). For the vesicle corresponding to Fig.~\ref{111102}, the best match between theory and experiments, i.e., the lowest chi-square, was obtained for $\sigma_0=(1.5\pm0.5) \times 10^{-8}$~N/m, and for the vesicle corresponding to Fig.~\ref{111024}, it was obtained for $\sigma_0=(5\pm1) \times 10^{-7}$~N/m. These values are in the expected range, and the data corresponding to the most flaccid vesicle is best matched by the lowest tension, which is satisfactory.

The experimental data presented in Figs.~\ref{111102} and \ref{111024} is slightly noisy. This is due to the fact that we have focused on small deformations, of 1 to 5~$\mu$m, in order to remain in the domain of application of our linear theoretical description. For such small deformations, all sources of noise (e.g., vibrations) become important, and the pixel size also becomes limiting. Our image treatment plugins were equipped with subpixel resolution in order to improve this point (see, e.g., Ref.~\cite{Crocker95}). Besides, it can be noted that the injection phase is more noisy than the relaxation phase, especially on Fig.~\ref{111024}. This is due to fluctuations of the injection pressure, which seem to occur mostly at the beginning of the injection phase. In particular, the excessive overshoots observed at the beginning of the injection on Fig.~\ref{111024} are very likely due to this artifact.

Figs.~\ref{111102} and \ref{111024} show a good agreement between our experimental data and the results of our theoretical description. In particular, our theory predicts the right timescales of deformation and of relaxation, and also the right variation of these timescales with the distance $z_0$ between the membrane and the micropipette. The increase of the timescales with $z_0$ comes from two different factors. First, the diffusion of the HO$^-$ ions takes longer if $z_0$ is larger. Second, when $z_0$ increases, the width of the modified membrane zone that deforms increases, so that smaller wavevectors have a higher importance in $\hat h (q,t)$. As the relaxation timescales of the membrane, which correspond to the inverse of the eigenvalues of $M$, all increase when $q$ decreases (see Eq.~(\ref{ED_defs})), this yields longer timescales. 

Besides, the timescales involved are shorter if the vesicle is more tense, for instance they are shorter in Fig.~\ref{111024} than in Fig.~\ref{111102}. This can be understood as follows: one of the relaxation timescales of the membrane, which corresponds to the inverse of one of the eigenvalues of $M$ (see Eq.~(\ref{ED_defs})), can be approximated by $4\eta/\sigma_0 q$ for the wavevectors $q$ with largest weight in $\hat h (q)$. This timescale decreases when $\sigma_0$ increases. More qualitatively, a tense membrane will tend to relax faster once it has been deformed.

While we have focused our discussion on the height $H(t)=h(r=0,t)$ of the membrane deformation in front of the micropipette, the full deformation profiles $h(r,t)$ are available both experimentally and theoretically in our work. We observe qualitative agreement between theory and experiment regarding the spatial profile of $h$ (and $\phi$), but the effect of vesicle curvature makes it difficult to push further the quantitative analysis away from the micropipette axis.

\section*{Conclusion}

We have studied experimentally and theoretically the deformation of a biomimetic lipid membrane in response to a local pH increase obtained by microinjecting a basic solution close to the membrane. Experimentally, we have measured the deformation height during and after the injection, and we have observed directly the pH profile on the membrane at the same time, using a pH-sensitive fluorescent membrane marker. Theoretically, our description of the phenomenon takes into account the linear dynamics of the membrane, and it also fully accounts for the time-dependent profile of the fraction of chemically modified lipids in the membrane. This profile results from the diffusion of the basic solution in the water that surrounds the membrane during and after the microinjection. We have compared experimental data regarding the height of the deformation to the results of our theoretical description, in the regime of small deformations, and we have obtained a good agreement between theory and experiments. 

Experimentally, it would be interesting to measure the vesicle tension through a micropipette~\cite{Evans90} at the same time as the microinjection is performed. It would perhaps become possible to adjust the intermonolayer friction coefficient $b$ (and hence to measure it) if the tension was known precisely. However, since high tensions yield shorter timescales and smaller deformations, it would be necessary to control precisely very small tensions through the micropipette.

From a theoretical point of view, our description is general and applies to a local injection of any substance that reacts reversibly with the membrane lipids. It could be improved by taking into account the curvature of the vesicle instead of taking a flat membrane as a reference state. Besides, it would be interesting to include nonlinear effects in order to describe larger, more dramatic, deformations, such as tubulation~\cite{Fournier09}.

The study of the response of a membrane to a local modification of its environment is a promising field. From the point of view of membrane physics, studying the spatiotemporal response of a membrane to a local modification can give access to membrane properties that were not accessible before. For instance, the ratio of the spontaneous curvature change to the equilibrium density change caused by a chemical modification cannot be determined from an analysis of static and global modifications~\cite{Lee99}. In the theoretical work Ref.~\cite{Futur}, we present a way of determining this ratio from the study of the dynamical response of a membrane to a continuous local injection. Hence, an interesting perspective would be to study continuous injections experimentally, which could yield an experimental measurement of this ratio. However, the injection pressure would have to be very stable for this to be done satisfactorily. More generally, we hope that studying the response of a biomimetic membrane to a local modification of its environment will help to shed light onto the relation between cellular phenomena and small-scale environment changes.

\section*{Acknowledgements}
We thank R\'emy Colin for his precious help with writing the image treatment plugins.

We thank the JSPS Core-to-Core Program ``International research network for non-equilibrium dynamics of soft matter'' for supporting this work.

\footnotesize{\providecommand*{\mcitethebibliography}{\thebibliography}
\csname @ifundefined\endcsname{endmcitethebibliography}
{\let\endmcitethebibliography\endthebibliography}{}

}

\end{document}